\documentclass[twocolumn,showpacs,preprintnumbers,superscriptaddress]{revtex4}%
\usepackage{amssymb}
\usepackage{amsmath}
\usepackage{graphicx}
\usepackage{epstopdf}
\usepackage{dcolumn}
\usepackage{bm}
\usepackage{amsfonts}%
\providecommand{\U}[1]{\protect\rule{.1in}{.1in}}
\providecommand{\U}[1]{\protect\rule{.1in}{.1in}}
\begin{document}
	\title{Mode locking in an optomechanical cavity}
	\author{Eyal Buks}
	\affiliation{Andrew and Erna Viterbi Department of Electrical Engineering, Technion, Haifa
		32000 Israel}
	\author{Roei Levi}
	\affiliation{Andrew and Erna Viterbi Department of Electrical Engineering, Technion, Haifa
		32000 Israel}
	\author{Ivar Martin}
	\affiliation{Materials Science Division, Argonne National Laboratory, Argonne, Illinois
		60439, USA}
	\date{\today }
	
	\begin{abstract}
		We experimentally study a fiber-based optical ring cavity integrated with a
		mechanical resonator mirror and an optical amplifier. The device exhibits a
		variety of intriguing nonlinear effects including synchronization and
		self-excited oscillation. Passively generated optical pulses are observed when
		the frequency of the optical ring cavity is tuned very close to the mechanical
		frequency of the suspended mirror. The optical power at the threshold of this
		process of mechanical mode locking is found to be related to quantum noise of
		the optical amplifier.
		
	\end{abstract}
	\maketitle





	Optomechanical cavities \cite{Braginsky_653,
		Hane_179,Gigan_67,Metzger_1002,Kippenberg_1172,Favero_104101,Marquardt2009,Poot_273}
	are widely employed for various sensing applications. The effect of radiation
	pressure typically governs the optomechanical coupling (i.e. the coupling
	between the electromagnetic cavity and the mechanical resonator that serves as
	a movable mirror) when the finesse of the optical cavity is sufficiently high,
	whereas, bolometric effects can contribute to the optomechanical coupling when
	optical absorption by the vibrating mirror is significant \cite{Metzger_1002,
		Jourdan_et_al_08,Marino&Marin2011PRE, Metzger_133903, Restrepo_860,
		Liberato_et_al_10,Marquardt_103901,
		Paternostro_et_al_06,Yuvaraj_430,Zaitsev_046605}.%

	\begin{figure}
		[ptb]
		\begin{center}
			\includegraphics[
			height=2.1106in,
			width=1.6337in
			]%
			{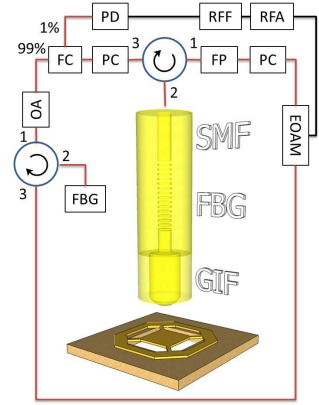}%
			\caption{The experimental setup. Red lines represent single mode optical
				fibers, whereas black lines represent radio frequency transmission lines
				(coaxial cables). The lump elements integrated into the ring cavity are (in
				counter clockwise direction, starting from the top circulator): manual
				polarization controller (PC), fiber coupler (FC), optical amplifier (OA),
				circulator connected to a fiber Bragg grating (FBG), electro-optical amplitude
				modulator (EOAM), electrically controlled PC and a fiber polarizer (FP). The
				1\% output port of the FC is connected to a photodetector (PD). The output PD
				signal is connected to a radio frequency filter (RFF) and a radio frequency
				amplifier (RFA), and the RFA output signal can be used as a feedback signal
				feeding the EOAM. The single mode fiber is terminated near the mechanical
				mirror by an FBG (having a negligible effect, since its filtering band does
				not overlap the one of the other FBG, that is integrated on the left arm of
				the ORC) and a graded index fiber (GIF) serving as a lens with a focal
				distance of $40\operatorname{\mu m}$. Details of the fabrication process used
				for preparing the suspended $100\operatorname{\mu m}\times
				100\operatorname{\mu m}$ mechanical mirror made of aluminum can be found in
				\cite{Suchoi_033818}. All measurements are performed at temperature of
				$77\operatorname{K}$ and pressure well below $2\times10^{-5}%
				\operatorname{mbar}$.}%
			\label{Fig_SU}%
		\end{center}
	\end{figure}

	Here we study laser mode locking \cite{Haus_1173} in an optomechanical cavity.
	Some applications for optomechanical cavities with integrated gain medium have
	been proposed before, including cooling \cite{Huang_013821}, squeezing of
	noise \cite{Agarwal_043844,Hu_023807,Lu_093602}, controlling dynamical
	instabilities \cite{Pina_033835} and normal mode splitting \cite{Huang_033807}%
	. Moreover, optomechanical cavities driven by externally injected pulses have
	been studied in \cite{Vanner_16182,Braginskii_27,Muhonen_1812_09720}. Optical
	pulses generated by externally modulating cavity length have been discussed in
	\cite{Henneberger_2189,Smith_51,Baranov_2123}. In the current study we explore
	a new method of passive mode locking, which is based on bolometric
	optomechanical coupling.
	
	The optical ring cavity (ORC) is schematically depicted by Fig. \ref{Fig_SU}.
	Optical gain is generated by a C-band Erbium-doped fiber optical amplifier
	(OA), and loss in the ORC is controlled by adjusting the direct voltage
	applied to the electro-optical amplitude modulator (EOAM). The state of
	polarization is controlled using two polarization controllers (PCs) and a
	fiber polarizer (FP). A fiber Bragg grating (FBG) \cite{Hill_1263} provides
	optical filtering. The optical signal is detected by a photodetector (PD)
	connected to a radio frequency filter (RFF) and a radio frequency amplifier
	(RFA), which can generate a feedback signal applied to the radio frequency
	port of the EOAM. Note that feedback signal is applied to the EOAM only for
	performing the measurements presented in Fig. \ref{Fig_ML_Synch} below.
	
	The effect of modulation (either by the moving mirror or by the EOAM) on the
	state of the ORC is discussed below. In the limit of small modulation
	amplitude the effect of noise cannot be disregarded, as was shown in
	\cite{Gordon_103901}. The ORC optical intensity $L_{\mathrm{H}}\left(
	t\right)  $ is expressed as $L_{\mathrm{H}}\left(  t\right)  =L_{0}%
	\mathcal{V}\left(  \omega_{\mathrm{R}}t\right)  $, where $L_{0}$ is the
	average intensity, $\omega_{\mathrm{R}}$ is the spacing between angular
	frequencies of the ORC modes, $t$ is time and the function $\mathcal{V}\left(
	s\right)  $ is expressed as a sum over all contributing modes of the ORC
	$\mathcal{V}\left(  s\right)  =N_{\mathrm{R}}^{-1}\left\langle \left\vert
	\sum_{m}r_{m}e^{i\left(  ms+\theta_{m}\right)  }\right\vert ^{2}\right\rangle
	$, where $N_{\mathrm{R}}$ is the number of ORC modes within the FBG filtering
	band, and the positive $r_{m}$ and the real $\theta_{m}$ are the amplitude and
	phase, respectively, of the $m$'th ORC mode. Consider an amplitude modulation
	applied to the ORC. When fluctuations in modes' amplitudes $r_{m}$ can be
	disregarded the evolution of the phases $\theta_{m}$ is governed by a set of
	coupled Langevin equations given by \cite{Gordon_103901}%
	\begin{equation}
	\dot{\theta}_{m}=\mu_{\mathrm{M}}\left(  \sin\left(  \theta_{m-1}-\theta
	_{m}\right)  +\sin\left(  \theta_{m+1}-\theta_{m}\right)  \right)
	+q_{m}\;,\label{theta_m dot}%
	\end{equation}
	where the terms proportional to the modulation amplitude $\mu_{\mathrm{M}}$
	represent the contribution of modulation-generated sidebands of neighboring
	modes, and the terms $q_{m}$ represent white noise satisfying correlation
	relations given by $\left\langle q_{m^{\prime}}\left(  t^{\prime}\right)
	q_{m^{\prime\prime}}^{\ast}\left(  t^{\prime\prime}\right)  \right\rangle
	=2T_{\mathrm{N}}\delta_{m^{\prime}m^{\prime\prime}}\delta\left(  t^{\prime
	}-t^{\prime\prime}\right)  $, where $T_{\mathrm{N}}$ is a constant. The main
	source of noise in the current experiment is quantum noise of the OA
	\cite{Giles_271}. In term of the Hamiltonian $\mathcal{H}\left(  \left\{
	\theta_{m}\right\}  \right)  $, which is given by $\mathcal{H}=-\mu
	_{\mathrm{M}}\sum_{m}\cos\left(  \theta_{m-1}-\theta_{m}\right)  $, Eq.
	(\ref{theta_m dot}) can be expressed as $\dot{\theta}_{m}=-\partial
	\mathcal{H}/\partial\theta_{m}+q_{m}$. In steady state the probability
	distribution $\mathcal{P}\left(  \left\{  \theta_{m}\right\}  \right)  $ is
	given by $\mathcal{P}=Z^{-1}e^{-\mathcal{H}/T_{\mathrm{N}}}$, where $Z$ is the
	partition function \cite{Risken_Fokker-Planck}. In the limit of weak noise the
	following holds $\left\langle \left(  \theta_{m-1}-\theta_{m}\right)
	^{2}\right\rangle =2\beta_{\mathrm{N}}$, where $\beta_{\mathrm{N}%
	}=T_{\mathrm{N}}/2\mu_{\mathrm{M}}$, and thus the phase correlation function
	is given by $\left\langle e^{i\left(  \theta_{m-k}-\theta_{m}\right)  ^{2}%
	}\right\rangle =e^{-\left\vert k\right\vert \beta_{\mathrm{N}}}$. Using these
	results one finds that $\mathcal{V}\left(  s\right)  =\mathcal{T}%
	_{\beta_{\mathrm{N}}}\left(  s\right)  $ in the limit of weak modulation
	amplitude $N_{\mathrm{R}}^{-1}\ll\beta_{\mathrm{N}}$, where the nicknamed comb
	function $\mathcal{T}_{\beta}\left(  s\right)  $, which is given by%
	\begin{equation}
	\mathcal{T}_{\beta}\left(  s\right)  =\sum\limits_{k=-\infty}^{\infty
	}e^{iks-\left\vert k\right\vert \beta}=\frac{\sinh\beta}{\cosh\beta-\cos
		s}\;,\label{T_beta(s)}%
	\end{equation}
	represents a periodic train of pulses having linewidth given by $\beta
	/2+O\left(  \beta^{2}\right)  $, and the averaged value of $\mathcal{T}%
	_{\beta}\left(  s\right)  $ is unity for any given $\beta$.
	
	In the other extreme of relatively large modulation amplitude $N_{\mathrm{R}%
	}\beta_{\mathrm{N}}\ll1$ optical noise can be approximately disregarded.
	Consider a Gaussian optical pulse \cite{Kuizenga_694} having amplitude
	$E\left(  t\right)  =E_{0}\exp\left(  -\gamma t^{2}+i\omega_{\mathrm{p}%
	}t\right)  $ circulating inside the ORC, where $E_{0}$ is a complex constant,
	$\gamma=\gamma^{\prime}+i\gamma^{\prime\prime}$, $\gamma^{\prime
	}=\operatorname{Re}\gamma>0$ determines the width of the pulse, $\gamma
	^{\prime\prime}=\operatorname{Im}\gamma$ represents a linear chirp, the real
	$\omega_{\mathrm{p}}$ is the optical angular frequency and $t$ is time. The
	effect of each of the lump elements integrated into the ORC on the mode shape
	is characterized by either a time-like or a frequency-like M\"{o}bius
	transformation $\gamma_{\mathrm{out}}^{-1}=\left(  A\gamma_{\mathrm{in}}%
	^{-1}+B\right)  /\left(  C\gamma_{\mathrm{in}}^{-1}+D\right)  $
	\cite{Nakazawa_1075}. For a frequency-like transformation characterized by the
	parameter $\gamma_{\mathrm{F}}$ one has $\gamma_{\mathrm{out}}^{-1}%
	=\gamma_{\mathrm{in}}^{-1}+\gamma_{\mathrm{F}}^{-1}$, i.e. $B=\gamma
	_{\mathrm{F}}^{-1}$ and $C=0$, whereas for a time-like transformation
	characterized by the parameter $\gamma_{\mathrm{T}}$ one has $\gamma
	_{\mathrm{out}}=\gamma_{\mathrm{in}}+\gamma_{\mathrm{T}}$, i.e. $B=0$ and
	$C=\gamma_{\mathrm{T}}$. For both cases $A=D=1$. Concatenating a time-like
	transformation with parameter $\gamma_{\mathrm{T}}$, and a frequency-like
	transformation with parameter $\gamma_{\mathrm{F}}$ yields a M\"{o}bius
	transformation with coefficients $A=1+g_{\mathrm{m}}^{2}$, $B=\gamma
	_{\mathrm{F}}^{-1}$, $C=\gamma_{\mathrm{F}}g_{\mathrm{m}}^{2}$ and $D=1$,
	where $g_{\mathrm{m}}=\left(  \gamma_{\mathrm{T}}\gamma_{\mathrm{F}}%
	^{-1}\right)  ^{1/2}$.
	
	Two cases of mode locking are discussed below, one is based on the EOAM and
	the other on the moving mirror. The effect of both elements is characterized
	by a time-like transformation with a parameter $\gamma_{\mathrm{T}}$. Both OA
	and FBG optical filter give rise to a frequency-like transformation. The
	magnitude $\left\vert \gamma_{\mathrm{F}}\right\vert $ of the parameter
	$\gamma_{\mathrm{F}}$ can be expressed in terms of an effective optical
	wavelength band $\Delta\lambda$ using the relation $\left\vert \gamma
	_{\mathrm{F}}\right\vert ^{1/2}=\left(  \left(  2\pi c\right)  /\left(
	\lambda_{\mathrm{L}}n_{\mathrm{eff}}\right)  \right)  \left(  \Delta
	\lambda/\lambda_{\mathrm{L}}\right)  $, where $c$ is the speed of light in
	vacuum, $n_{\mathrm{eff}}=1.47$ is the fiber mode effective refractive index
	and $\lambda_{\mathrm{L}}=1550%
	\operatorname{nm}%
	$ is the optical wavelength. For the OA in the current experiment
	$\Delta\lambda\simeq50%
	\operatorname{nm}%
	$, whereas for the FBG $\Delta\lambda\simeq0.2%
	\operatorname{nm}%
	$, and thus to a good approximation the transformation parameter
	$\gamma_{\mathrm{F}}$ can be evaluated by disregarding the effect of the OA.
	For both methods of mode locking that are employed in the current experiment
	the dimensionless parameter $\left\vert g_{\mathrm{m}}\right\vert $ is at most
	about $10^{-5}$. When $\left\vert g_{\mathrm{m}}\right\vert \ll1$ it is
	convenient to represent the discrete M\"{o}bius transformation by a continuous
	differential equation of the normalized pulse parameter $g=\gamma
	/\gamma_{\mathrm{F}}$. To lowest nonvanishing order in $g$ and $g_{\mathrm{m}%
	}$ one finds that $g$ evolves according to $\mathrm{d}g/\mathrm{d}%
	\tau_{\mathrm{R}}=g_{\mathrm{m}}^{2}-g^{2}$, where $\tau_{\mathrm{R}%
	}=t/t_{\mathrm{R}}$, and $t_{\mathrm{R}}$ is the ORC period time. Out of the
	two fixed points $\pm g_{\mathrm{m}}$, only the one having a positive real
	value is stable. The solution is given by $g=g_{\mathrm{m}}\tanh\left(
	g_{\mathrm{m}}\left(  \tau_{\mathrm{R}}-\tau_{\mathrm{R}0}\right)  \right)  $,
	where $\tau_{\mathrm{R}0}$ is a constant.
	
	Mode locking based on the EOAM is explored without integrating the mechanical
	mirror. This is done by placing the optical fiber (which is connected to a
	piezoelectric positioner) above a pad near the trampoline. The pad, which is
	made of the same aluminum and silicon nitride layers (as the suspended mirror), serves as a static
	mirror. In this mode of operation active mode locking can be induced by
	driving the EOAM at a frequency $\omega_{\mathrm{AM}}/2\pi$ close to the ORC
	frequency $\omega_{\mathrm{R}}/2\pi=371.3%
	\operatorname{kHz}%
	$. In addition pulses can be obtained by the method of regenerative mode
	locking (RGML) \cite{Huggett_186,Levy_148}, in which the PD signal generates a
	feedback signal driving the EOAM (see Fig. \ref{Fig_SU}). The phase of the
	pulses generated by RGML can be locked by simultaneously driving the EOAM at a
	frequency $\omega_{\mathrm{AM}}/2\pi$ close to the ORC frequency
	$\omega_{\mathrm{R}}/2\pi$. This is demonstrated in Fig. \ref{Fig_ML_Synch}%
	(a), which shows a color-coded plot of the spectral density (measured using a
	spectrum analyzer) of the PD signal as a function of the voltage amplitude
	$V_{\mathrm{AM}}$ of a modulation signal at a fixed frequency of
	$\omega_{\mathrm{AM}}/2\pi=371.4%
	\operatorname{kHz}%
	=\omega_{\mathrm{R}}/2\pi+100%
	\operatorname{Hz}%
	$ applied to the EOAM. Synchronization occurs in the region $V_{\mathrm{AM}%
	}\geq V_{\mathrm{AM},0}=0.156%
	\operatorname{V}%
	$.  In the unlocked region, multiple side bands emerge via a continuous transition. Such side band structure is consistent with the pulse remaining synchronized with the modulation for extended periods of time, interlaced by rapid ``phase slip" events.  Unlocking  can be well approximated  by solving the
	equation of motion for the relative phase $\varphi_{\mathrm{S}}$ between the
	pulsing oscillation and the applied modulation,  given by
	\cite{Strogatz_Nonlinear,Buks_032202}%
	\begin{equation}
	\frac{\mathrm{d}\varphi_{\mathrm{S}}}{\mathrm{d}\tau}+\sin\varphi_{\mathrm{S}%
	}=i_{\mathrm{b}}\;,\label{varphi eom}%
	\end{equation}
	where $i_{\mathrm{b}}=\left(  \omega_{\mathrm{AM}}-\omega_{\mathrm{R}}\right)
	/\left(  \zeta_{\mathrm{AM}}V_{\mathrm{AM}}\right)  $ is a normalized
	detuning, the coefficient $\zeta_{\mathrm{AM}}$ is given by $\zeta
	_{\mathrm{AM}}=\left(  \omega_{\mathrm{AM}}-\omega_{\mathrm{R}}\right)
	/V_{\mathrm{AM},0}$ (i.e. $i_{\mathrm{b}}=1$ at the onset of synchronization)
	and $\tau=\zeta_{\mathrm{AM}}V_{\mathrm{AM}}t$ is a dimensionless time
	variable. The solution of Eq. (\ref{varphi eom}) in the region $\left\vert
	i_{\mathrm{b}}\right\vert >1$ can be expressed in terms of the comb function
	$\mathcal{T}_{\beta}$ that is defined by Eq. (\ref{T_beta(s)}) as
	$\mathrm{d}\varphi_{\mathrm{S}}/\mathrm{d}\tau=\sinh\beta_{\mathrm{b}%
	}\mathcal{T}_{\beta_{\mathrm{b}}}\left(  \tau\sinh\beta_{\mathrm{b}}%
	+\theta_{\mathrm{b}}\right)  $, where $\beta_{\mathrm{b}}=\cosh^{-1}%
	i_{\mathrm{b}}$ and the phase $\theta_{\mathrm{b}}$ is given by $\theta
	_{\mathrm{b}}=\pi-\tan^{-1}\left(  \sinh\beta_{\mathrm{b}}\right)  $
	\cite{Buks_032202}. The resulting spectrum is shown in Fig. \ref{Fig_ML_Synch}(b).
	
\begin{figure}
		[ptb]
		\begin{center}
			\includegraphics[
			height=2.5949in,
			width=3.4537in
			]%
			{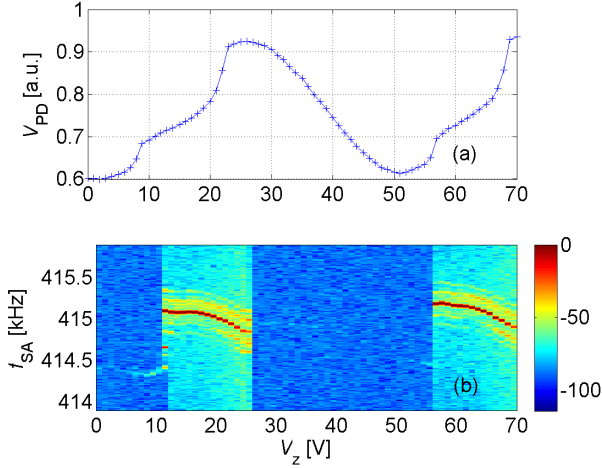}%
			\caption{Self-excited oscillation. (a) The averaged PD voltage $V_{\mathrm{PD}%
				}$ as a function of $V_{z}$. (b) The spectral density of the PD signal (in
				normalized dB units).}%
			\label{Fig_SEO}%
		\end{center}
	\end{figure}

\begin{figure}
		[ptb]
		\begin{center}
			\includegraphics[
			height=4.3098in,
			width=3.4509in
			]%
			{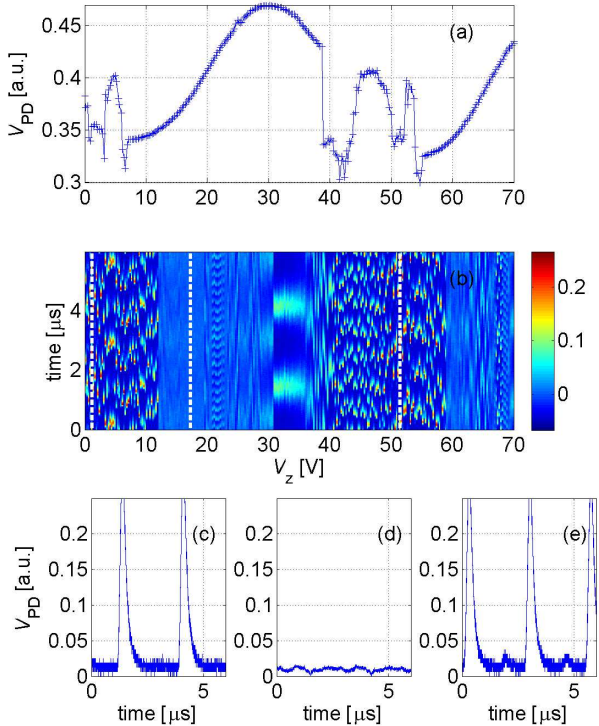}%
			\caption{Mechanical mode locking. (a) The averaged PD voltage $V_{\mathrm{PD}%
				}$ as a function of the voltage $V_{z}$ applied to the piezoelectric motor.
				(b) Oscilloscope time traces of the PD signal. The values of $V_{z}$
				corresponding to the time traces presented in (c), (d) and (e) are indicated
				by the three overlaid vertical white dotted lines in (b).}%
			\label{Fig_MML}%
		\end{center}
	\end{figure}

	The effect of the mechanical mirror is explored by studying two dynamical
	instabilities, self-excited oscillation (SEO) (see Fig. \ref{Fig_SEO}) and
	mechanical mode locking (MML) (see Fig. \ref{Fig_MML}). For studying SEO the
	ORC frequency $\omega_{\mathrm{R}}/2\pi=2.48%
	\operatorname{MHz}%
	$ is not tuned close to the mechanical frequency $\omega_{\mathrm{m}}/2\pi=415%
	\operatorname{kHz}%
	$, whereas the MML measurements are performed after adjusting the total length
	$L_{\mathrm{R}}$ of the ORC to satisfy the condition $\omega_{\mathrm{R}%
	}\simeq\omega_{\mathrm{m}}$. As was mentioned above, for both cases no
	feedback signal is applied to the EOAM.
	
	Let $L_{\mathrm{SC}}$ be the total length of the optical cavity that is formed
	between the fiber's tip and the mirror. This cavity is henceforth referred to
	as the short cavity (SC), to avoid confusion with the much longer fiber ORC.
	The length $L_{\mathrm{SC}}$ can be controlled by adjusting the voltage
	$V_{z}$ that is applied to one of the piezoelectric motors moving the fiber.
	The plot shown in Fig. \ref{Fig_SEO}(a) presents the measured averaged PD
	voltage $V_{\mathrm{PD}}$ as a function of $V_{z}$. The two local minima
	points of $V_{\mathrm{PD}}$ (obtained with $V_{z}=2%
	\operatorname{V}%
	$ and $V_{z}=51%
	\operatorname{V}%
	$, respectively) represent two optical resonances of the SC (i.e. the SC
	length $L_{\mathrm{SC}}$ is shortened by $\lambda_{\mathrm{L}}/2$ by
	increasing the voltage from the value $V_{z}=2%
	\operatorname{V}%
	$ to the value $V_{z}=51%
	\operatorname{V}%
	$). The spectral density of the PD signal (measured using a spectrum analyzer)
	is shown in Fig. \ref{Fig_SEO}(b). The intense spectral peak that is observed
	in the regions $V_{z}\in\left[  11%
	\operatorname{V}%
	,26%
	\operatorname{V}%
	\right]  $ and $V_{z}\in\left[  56%
	\operatorname{V}%
	,70%
	\operatorname{V}%
	\right]  $ occurs due to mechanical SEO.
	
	The response of the system is next explored in the region where $\omega
	_{\mathrm{R}}\simeq\omega_{\mathrm{m}}$ [$\omega_{\mathrm{R}}$ is tuned to
	this value by adjusting the length $L_{\mathrm{R}}$ of the ORC to the value
	$\left(  c/n_{\mathrm{eff}}\right)  /\left(  \omega_{\mathrm{m}}/2\pi\right)
	=553.88%
	\operatorname{m}%
	$, where $\omega_{\mathrm{m}}/2\pi=368.2%
	\operatorname{kHz}%
	$ for the mechanical mirror used for these measurements]. The accuracy of this
	procedure is typically about $0.01%
	\operatorname{Hz}%
	$. The measured averaged PD voltage $V_{\mathrm{PD}}$ as a function of $V_{z}$
	is shown in Fig. \ref{Fig_MML}(a). The color-coded plot in Fig. \ref{Fig_MML}%
	(b) shows time traces measured by an oscilloscope connected to the PD. The
	time traces presented in Fig. \ref{Fig_MML}(c-e) are obtained for the values
	$V_{z}=1.2%
	\operatorname{V}%
	$, $V_{z}=17.2%
	\operatorname{V}%
	$ and $V_{z}=51.5%
	\operatorname{V}%
	$, respectively [these values are indicated by overlaid white dotted lines in
	Fig. \ref{Fig_MML}(b)]. The pulses shown in Fig. \ref{Fig_MML} are attributed
	to mirror motion that effectively generates modulation at the frequency of the
	mechanical oscillation. From the measured linewidth of the pulses and the
	relation $g_{\mathrm{m}}=\left(  \gamma_{\mathrm{T}}\gamma_{\mathrm{F}}%
	^{-1}\right)  ^{1/2}$ one finds that the modulation amplitude $\left\vert
	\gamma_{\mathrm{T}}\right\vert ^{1/2}\simeq4%
	\operatorname{MHz}%
	$ for the narrowest peaks seen in Fig. \ref{Fig_MML}. Note that, contrary to
	the case of SEO that is observed with red detuning (see Fig \ref{Fig_SEO}),
	the MML shown in Fig. \ref{Fig_MML} is obtained mainly with blue detuning.%
	
	\begin{figure}
		[ptb]
		\begin{center}
			\includegraphics[
			height=2.5949in,
			width=3.4537in
			]%
			{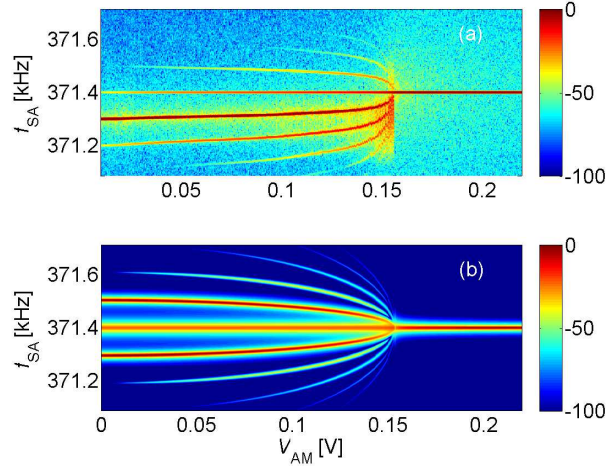}%
			\caption{Synchronization of RGML. (a) The spectral density of the PD signal as
				a function of the EOAM modulation amplitude $V_{\mathrm{AM}}$. (b) The
				calculated spectral density is obtained from the solution of Eq.
				(\ref{varphi eom}).}%
			\label{Fig_ML_Synch}%
		\end{center}
	\end{figure}

	Both effects of SEO and MML are attributed to bolometric optomechanical
	coupling. Consider an optical cavity with a movable mirror having mass
	$m_{\mathrm{m}}$, intrinsic mechanical angular frequency $\omega_{\mathrm{m}}$
	and an intrinsic mechanical damping rate $\gamma_{\mathrm{m}}\ll
	\omega_{\mathrm{m}}$. It is assumed that the angular resonance frequency of
	the mechanical resonator depends on the temperature $T_{\mathrm{m}}$ of the
	suspended mirror. For small deviation $T_{\mathrm{R}}=T_{\mathrm{m}%
	}-T_{\mathrm{b}}$ of $T_{\mathrm{m}}$ from the base temperature $T_{\mathrm{b}%
	}$ (i.e. the temperature of the supporting substrate) it is taken to be given
	by $\omega_{\mathrm{m}}+\Theta_{\mathrm{PH}}T_{\mathrm{R}}$, where
	$\Theta_{\mathrm{PH}}$ is a constant. Furthermore, to model the effect of
	thermal deformation \cite{Metzger_133903} it is assumed that a temperature
	dependent force given by $F_{\mathrm{T}}=\Theta_{\mathrm{FH}}T_{\mathrm{R}}$,
	where $\Theta_{\mathrm{FH}}$ is a constant, acts on the mechanical resonator.
	The mechanical oscillator's equation of motion is given by $\ddot
	{x}_{\mathrm{m}}+2\gamma_{\mathrm{m}}\dot{x}_{\mathrm{m}}+\left(
	\omega_{\mathrm{m}}+\Theta_{\mathrm{PH}}T_{\mathrm{R}}\right)  ^{2}%
	x_{\mathrm{m}}=m_{\mathrm{m}}^{-1}F_{\mathrm{T}}$, where an overdot denotes
	differentiation with respect to time. Intrinsic mechanical nonlinearities of
	the mirror are disregarded, i.e. it is assumed that nonlinear behavior
	exclusively originates from bolometric optomechanical coupling.%
	
	%
	
	The time evolution of the relative temperature $T_{\mathrm{R}}$ is governed by
	the thermal balance equation $\dot{T}_{\mathrm{R}}=H_{\mathrm{m}}%
	-\kappa_{\mathrm{m}}T_{\mathrm{R}}$, where $H_{\mathrm{m}}$ is proportional to
	the optically-induced heating power and $\kappa_{\mathrm{m}}\simeq
	0.01\omega_{\mathrm{m}}$ is the thermal decay rate. The relative phase between
	heating $H_{\mathrm{m}}$ and relative temperature $T_{\mathrm{R}}$ for a
	steady state solution of the thermal balance equation is $\theta_{\mathrm{T}%
	}-\pi/2$, where $\theta_{\mathrm{T}}=\tan^{-1}\kappa_{\mathrm{m}}%
	/\omega_{\mathrm{m}}$. The heating term $H_{\mathrm{m}}$ (which is
	proportional to the intra-cavity optical power incident on the suspended
	mirror) is expressed as $H_{\mathrm{m}}=L_{\mathrm{H}}\left(  t\right)
	A_{\mathrm{H}}\left(  x_{\mathrm{m}}\right)  $, where $L_{\mathrm{H}}$ is the
	optical intensity. To second order in the mechanical displacement
	$x_{\mathrm{m}}$ the optical absorption coefficient of the mirror
	$A_{\mathrm{H}}$ is expressed as $A_{\mathrm{H}}=A_{\mathrm{H}0}\left(
	1+k_{\mathrm{A}1}x_{\mathrm{m}}+k_{\mathrm{A}2}x_{\mathrm{m}}^{2}\right)
	+O\left(  x_{\mathrm{m}}^{3}\right)  $. The dependency of $A_{\mathrm{H}}$ on
	the mechanical displacement $x_{\mathrm{m}}$ originates from interference in
	the SC that is formed between the fiber's tip and the mechanical mirror (note
	that the length of the SC $\simeq40%
	\operatorname{\mu m}%
	$ is much shorter than the coherence length $\lambda_{\mathrm{L}}^{2}%
	/\Delta\lambda$, where $\Delta\lambda=0.2%
	\operatorname{nm}%
	$ is the filtering bandwidth of the FBG). Note that the finesse of the SC is
	by far sufficiently low to allow disregarding retardation in the response of
	the SC to mechanical displacement.%

	\begin{figure}
		[ptb]
		\begin{center}
			\includegraphics[
			height=1.1789in,
			width=3.4537in
			]%
			{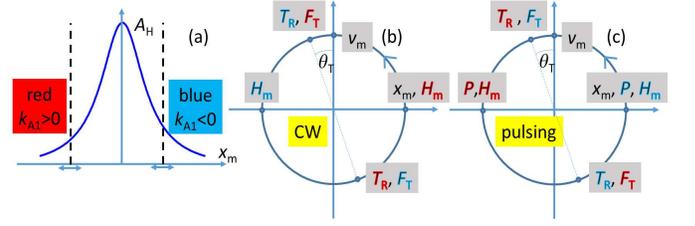}%
			\caption{Relative phase of the thermal force $F_{\mathrm{T}}$. (a) Sketch of
				the mirror optical absorption coefficient $A_{\mathrm{H}}$ as a function of
				mechanical displacement $x_{\mathrm{m}}$ near an optical resonance. The
				absorption coefficient $k_{\mathrm{A}1}$ is negative (positive) when the
				cavity is blue (red) detuned. In plots (b) and (c) the overlaid dots on the
				circles schematically represent the following oscillating variables:
				mechanical displacement $x_{\mathrm{m}}$, mechanical velocity $v_{\mathrm{m}}%
				$, optically-induced heating $H_{\mathrm{m}}$, pulsing $P$, relative
				temperature $T_{\mathrm{R}}$ and thermal force $F_{\mathrm{T}}$. The relative
				angles between dots represent the corresponding relative phase between the
				oscillating variables. The plot in (b) represents the contribution of
				continuous wave (CW) optical intensity, whereas the contribution of pulsing is
				described by the plot in (c). For both cases the relative phase between
				mechanical displacement $x_{\mathrm{m}}$ and heating $H_{\mathrm{m}}$ depends
				on the detuning of the short cavity. The assumed detuning is indicated by
				coloring the letters $H_{\mathrm{m}}$, $P$, $T_{\mathrm{R}}$ and
				$F_{\mathrm{T}}$ accordingly. Since $\Theta_{\mathrm{FH}}<0$ the thermal force
				$F_{\mathrm{T}}$ is out of phase with respect to the relative temperature
				$T_{\mathrm{R}}$.}%
			\label{Fig_RelPh}%
		\end{center}
	\end{figure}
	
	The effect of the bolometric coupling on the dynamics of the optomechanical
	cavity has been extensively studied before for the case where light with a
	constant intensity is externally injected \cite{Metzger_1002,
		Jourdan_et_al_08,Marino&Marin2011PRE, Metzger_133903, Restrepo_860,
		Liberato_et_al_10,Marquardt_103901,
		Paternostro_et_al_06,Yuvaraj_430,Zaitsev_046605}. Most results of this
	analysis are applicable for the SEO measurements shown in Fig. \ref{Fig_SEO},
	for which the ORC frequency $\omega_{\mathrm{R}}$ and the mechanical frequency
	$\omega_{\mathrm{m}}$ are incommensurable. On the other hand, when the
	detuning between the ORC and mechanical frequencies is sufficiently small, the
	motion-induced modulation may have a significant effect on the intra-cavity
	optical intensity $L_{\mathrm{H}}$, as is demonstrated by the MML measurements
	shown in Fig. \ref{Fig_MML}.
	
	In the limit of small mechanical displacement the main effect of the
	optomechanical coupling originates from two terms of $H_{\mathrm{m}%
	}=L_{\mathrm{H}}A_{\mathrm{H}}$ both oscillating at the mechanical frequency,
	the first one is due to motion-induced oscillation of the absorption
	$A_{\mathrm{H}}$, and the second one is due to motion-induced modulation in
	the optical intensity $L_{\mathrm{H}}\left(  t\right)  $ [see Eq.
	(\ref{T_beta(s)})]. The effect of both terms can be taken into account by
	replacing the mechanical angular frequency $\omega_{\mathrm{m}}$ and
	mechanical damping rate $\gamma_{\mathrm{m}}$ by effective values given by%
	\begin{align}
	\omega_{\mathrm{m},\mathrm{eff}}  &  =\omega_{\mathrm{m}}+\frac{\kappa
		_{\mathrm{m}}}{\omega_{\mathrm{m}}}\left(  \gamma_{\mathrm{H}0}+\gamma
	_{\mathrm{H}1}\right)  \;,\label{omega_m eff}\\
	\gamma_{\mathrm{m},\mathrm{eff}}  &  =\gamma_{\mathrm{m}}+\gamma_{\mathrm{H}%
		0}+\gamma_{\mathrm{H}1}\;. \label{gamma_m eff}%
	\end{align}
	In both regions of SEO and MML the damping rate $\gamma_{\mathrm{m}%
		,\mathrm{eff}}$ becomes negative, and consequently the system becomes
	unstable. Nonlinear corrections to Eqs. (\ref{omega_m eff}) and
	(\ref{gamma_m eff}) can be evaluated by taking into account both the
	parametric term proportional to $\Theta_{\mathrm{PH}}$ and the second order
	absorption term proportional to $k_{\mathrm{A}2}$ \cite{Zaitsev_1589}. Note,
	however, that these nonlinear terms are not needed for determining the
	threshold of both SEO and MML. The term $\gamma_{\mathrm{H}0}=k_{\mathrm{A}%
		1}\Theta_{\mathrm{FH}}L_{0}A_{\mathrm{H}0}/\left(  2m_{\mathrm{m}}%
	\omega_{\mathrm{m}}^{2}\right)  $ represents the contribution of the average
	value of the optical intensity $L_{0}$ \cite{Zaitsev_1589}, whereas the
	contribution of the optical intensity $L_{\mathrm{H}}$ component oscillating
	at the mechanical frequency is represented by the term $\gamma_{\mathrm{H}1}$.
	When the detuning between the ORC and mechanical frequencies is negligibly
	small $\gamma_{\mathrm{H}1}$ is found to be given by $\gamma_{\mathrm{H}%
		1}=-\left(  2\omega_{\mathrm{m}}/T_{\mathrm{N}}\right)  \gamma_{\mathrm{H}0}$
	[see Eq. (\ref{T_beta(s)})], whereas the term $\gamma_{\mathrm{H}1}$ can be
	disregarded when $\omega_{\mathrm{R}}$ and $\omega_{\mathrm{m}}$ are incommensurate.
	
	The dominant contribution to the effective noise parameter $T_{\mathrm{N}}$
	originates from quantum noise of the OA, which has a noise figure
	$\alpha_{\mathrm{NF}}$ given by $\alpha_{\mathrm{NF}}=2n_{\mathrm{PI}}\left(
	G_{\mathrm{OA}}-1\right)  /G_{\mathrm{OA}}=2.5$, where $G_{\mathrm{OA}}=1600$
	is the small signal gain and $n_{\mathrm{PI}}=1.25$ is the population
	inversion parameter \cite{Giles_271}. When thermal occupation of the optical
	modes is negligibly small the effective noise parameter $T_{\mathrm{N}}$ is
	given by $T_{\mathrm{N}}\simeq\left(  \gamma_{\mathrm{OM}}\alpha_{\mathrm{NF}%
	}G_{\mathrm{OA}}\right)  /\left(  4\left\langle n_{\mathrm{p}}\right\rangle
	\right)  $, where $\gamma_{\mathrm{OM}}\simeq0.1\times\omega_{\mathrm{R}}$ is
	a typical mode damping rate (dominated by both insertion loss of the EOAM and
	radiation loss of the SC), and where $\left\langle n_{\mathrm{p}}\right\rangle
	\simeq2\times10^{6}$ is the averaged photon number per mode for the
	measurements presented in Fig. \ref{Fig_MML}, and thus for this case
	$\left\vert \gamma_{\mathrm{H}1}/\gamma_{\mathrm{H}0}\right\vert
	=2\omega_{\mathrm{m}}/T_{\mathrm{N}}\simeq4\times10^{4}$ (note that
	$\left\langle n_{\mathrm{p}}\right\rangle $ is related to the power
	$P_{\mathrm{OA}}$ delivered by the OA by $P_{\mathrm{OA}}=\gamma_{\mathrm{OM}%
	}\hbar\omega_{\mathrm{p}}N_{\mathrm{R}}\left\langle n_{\mathrm{p}%
	}\right\rangle $, where $N_{\mathrm{R}}=L_{\mathrm{R}}\Delta\lambda
	/\lambda_{\mathrm{L}}^{2}$). The fact that $\left\vert \gamma_{\mathrm{H}%
		1}/\gamma_{\mathrm{H}0}\right\vert \gg1$ is demonstrated by the experimental
	observations that MML occurs in almost the entire region of blue detuning (see
	Fig. \ref{Fig_MML}), whereas for the same optical gain SEO occurs only in a
	partial region of red detuning (see Fig. \ref{Fig_SEO}).
	
	Note that the sign of the term $\gamma_{\mathrm{H}1}$ is opposite to the sign
	of $\gamma_{\mathrm{H}0}$. As is explained in the caption of Fig.
	\ref{Fig_RelPh}, this can be attributed to the fact that pulses generated by
	mode locking hit the mirror when the displacement-dependent optical absorption
	$A_{\mathrm{H}}\left(  x_{\mathrm{m}}\right)  $ obtains its minimum value.
	Both added damping rates $\gamma_{\mathrm{H}0}$ and $\gamma_{\mathrm{H}1}$
	[see Eq. (\ref{gamma_m eff})] depend on the detuning of the SC. When the SC is
	blue (red) detuned the absorption coefficient $k_{\mathrm{A}1}$ in Eq.
	(\ref{gamma_m eff}) is negative (positive) \cite{Baskin_563} (note that it is
	assumed that positive displacement $x_{\mathrm{m}}$ is in the outwards
	direction and that the cavity length $L_{\mathrm{SC}}$ is decreased when the
	piezoelectric motor voltage $V_{z}$ is increased). Aluminum has a thermal
	expansion coefficient higher than both silicon and silicon-nitride, and
	consequently it is expected that $\Theta_{\mathrm{FH}}<0$ and $\Theta
	_{\mathrm{PH}}<0$. Thus, for the device under study here it is expected that
	$\gamma_{\mathrm{H}0}<0$ when the SC is red detuned [see Eq.
	(\ref{gamma_m eff})]. This behavior is demonstrated by the SEO shown in Fig.
	\ref{Fig_SEO}. On the other hand, $\gamma_{\mathrm{H}1}<0$ when the SC is blue
	detuned. This is consistent with the experimental observation that MML occurs
	mainly with blue detuning (see Fig. \ref{Fig_MML}).
	
	In summary, we find that mode locking can be obtained by integrating gain
	medium into an optomechanical cavity. The threshold optical power for MML is
	found to be significantly lower than the corresponding value for SEO. Future
	study will explore applications of MML for sensing. For example, Braginsky has
	proposed a device called a speed meter, which allows monitoring a classical
	force acting on a mechanical resonator with sensitivity that can exceed the
	so-called standard quantum limit \cite{Braginsky_251,Braginskii_27}. In the
	steady state of MML the pulses hit the mirror when its velocity nearly
	vanishes, and thus the off reflected pulses mainly carry information about the
	velocity (rather than the position) of the mirror, therefore a sensor based on
	MML may serve as a sensitive speed meter.
	
	We thank Moshe Horowitz and Baruch Fischer for useful discussion.
	Argonne National Laboratory’s contribution is based upon work supported by Laboratory Directed Research and Development (LDRD) funding from Argonne National Laboratory, provided by the
	Director, Office of Science, of the U.S. Department of
	Energy under Contract No. DE-AC02-06CH11357.

	\bibliographystyle{ieeepes}
	\bibliography{acompat,Eyal_Bib}

\end{document}